\documentclass{article}

\usepackage{arxiv}

\usepackage[utf8]{inputenc} % allow utf-8 input
\usepackage[T1]{fontenc}    % use 8-bit T1 fonts
\usepackage{hyperref}       % hyperlinks
\usepackage{url}            % simple URL typesetting
\usepackage{booktabs}       % professional-quality tables
\usepackage{amsmath}
\usepackage{amssymb}
\usepackage{amsfonts}       % blackboard math symbols
\usepackage{nicefrac}       % compact symbols for 1/2, etc.
\usepackage{microtype}      % microtypography
\usepackage{lipsum}
\usepackage{graphicx}

\graphicspath{ {./images/} }

\title{Modeling the position and velocity
distribution of space objects by maximizing
entropy with energy constraint}

\author{
 Partha Chowdhury \\
 PhD Scholar \\
  Department of ECE\\
  Indraprastha Institute of Information Technology, Delhi\\
  New Delhi, 110020 \\
  \texttt{parthac@iiitd.ac.in} \\
   \And
 Sanat K Biswas \\
 Assistant Professor \\
  Department of ECE\\
  Indraprastha Institute of Information Technology, Delhi\\
  New Delhi, 110020 \\
  \texttt{sanat@iiitd.ac.in} \\
}

\begin{document}
\maketitle
\begin{abstract}
In this work, we have developed a 6-dimensional joint probability density function for the 3-dimensional position and 3-dimensional velocity vectors of space objects in the Low Earth Orbit (LEO) based on the Principle of Maximum Entropy (MaxEnt), adhering to the principle of energy conservation. For the problem under consideration, maximizing entropy subject to energy conservation ensures that the derived probability density function (PDF) is the best representation of the uncertainty of a space object while the sampled position and velocity vectors from the PDF adhere to the orbital dynamics. We approach the entropy maximization by constructing a Lagrangian functional incorporating the energy conservation constraint and the normalization constraint of the PDF using Lagrange multipliers, setting the functional derivative of the Lagrangian to zero. This PDF can be used to generate position and velocity samples for space objects without any prior assumption and can further be utilized for orbital uncertainty propagation either using the Monte-Carlo method or by direct propagation of the PDF through the Fokker-Planck Equation.
\end{abstract}

% keywords can be removed

\section{Introduction}
The exponential growth of satellite deployments in LEO has led to a significant rise in the risk of collisions in space. As of today, there are approximately 11,050 active satellites, with the majority operating in LEO. In addition to active satellites, the orbital environment is cluttered with millions of debris fragments, including those smaller than 10 cm in diameter, which pose a substantial threat to operational spacecraft. These small debris are difficult to track accurately, making collision avoidance a complex challenge.
Predicting the probabilistic distribution of space objects—including both satellites and debris—is critical for ensuring safe orbital operations. However, the uncertainties in their position and velocity make traditional deterministic tracking methods inadequate. Typically, Gaussian distributions are assumed to model these uncertainties, but they often fail accurately to capture the highly nonlinear and dynamic nature of orbital motion, which is governed by orbital mechanics and perturbative forces.
To overcome this limitation, we propose a novel MaxEnt approach to derive a joint 6D PDF for the 3D position and 3D velocity of space objects in LEO. MaxEnt states that when estimating a probability distribution with incomplete information, one should select the distribution that maximizes entropy while satisfying some constraints. In our case, we maximize entropy subject to energy conservation, ensuring that the derived PDF accurately represents the uncertainty in a physically consistent manner. Energy conservation, a key concept in classical and orbital mechanics, states that in a space object in orbit, the total specific mechanical energy—defined as the sum of kinetic energy and potential energy remains constant unless acted upon by external forces such as atmospheric drag, solar radiation pressure, or gravitational perturbations. By applying energy conservation in a PDF, we align the model with fundamental physical laws, improving its physical relevance and predictive accuracy.\\
We formulate this problem as a constrained optimization problem, where entropy maximization is performed using the Lagrangian method with energy conservation as a constraint. By solving the resulting functional derivative equation, we obtain an analytical expression for the probability distribution that is efficient enough to represent the orbital uncertainties. This MaxEnt-derived PDF allows us to generate position and velocity samples without making Gaussian assumptions, providing a more general framework for orbit uncertainty propagation. The resulting PDF can be propagated using Monte Carlo methods or directly through the Fokker-Planck equation, making it applicable to a wide range of space situational awareness problems.

One of the most widely used techniques in space object uncertainty representation is the Gaussian distribution assumption. The covariance-based uncertainty propagation model \cite{article} assumes that position and velocity uncertainties follow a normal distribution, allowing easy propagation using the Fokker-Planck Equation. However, this assumption breaks down in long-term propagation due to nonlinear perturbations (e.g., atmospheric drag, third-body effects) that introduce non-Gaussian behaviour \cite{article2}.

To improve accuracy, researchers have proposed higher-order Gaussian mixture models \cite{app13053069} or the Unscented Transform \cite{1271397} to approximate nonlinear effects. However, these methods still rely on Gaussian assumptions and do not fully account for the constraints imposed by the law orbital mechanics.

Due to the limitations of Gaussian models, researchers have explored non-Gaussian uncertainty representations. The Polynomial Chaos Expansion (PCE) method \cite{book1} and Monte Carlo sampling \cite{ROCHMAN2014367} allows for more flexible modelling of space object distributions. While these methods improve accuracy, they require significant computational resources and do not provide a systematic way to determine the most unbiased probability distribution.

The Fokker-Planck equation (FPE) has also been used to model the evolution of probability density functions over time \cite{inproceedings}. However, solving the FPE requires strong assumptions about force models and may not always yield closed-form solutions.

The Principle of Maximum Entropy (MaxEnt) provides a powerful framework for estimating probability distributions when only partial information is available. Initially introduced by Jaynes \cite{PhysRev.106.620}, MaxEnt has been applied in statistical mechanics, signal processing, and information theory \cite{kapur1990maximum}. In orbital mechanics, it has been explored for trajectory prediction but is not widely used for position-velocity uncertainty modelling.

Recent works have applied entropy-based methods to space situational awareness, such as the use of Rényi entropy for sensor fusion \cite{BARATPOUR20082544} and entropy minimization for orbit determination \cite{Demars2013EntropyBasedAF}. However, these works do not explicitly derive a joint position-velocity PDF under energy conservation constraints.

While previous methods have focused on Gaussian-based models, non-Gaussian numerical techniques, or entropy-based sensor fusion, our approach uniquely applies the MaxEnt principle to derive an analytical probability density function for space objects while explicitly enforcing energy conservation. This method ensures that the resulting distribution is the most unbiased representation of uncertainty while maintaining physical consistency with orbital mechanics.

\section{Problem Statement and Mathematical Formulation}
The primary objective of this work is to derive a 6D joint PDF for the position (r) and velocity (v) vectors of space objects in LEO, ensuring that the PDF satisfies the principle of maximum entropy and conserves the energy functional.
later in this work we will validate the derived PDF through numerical simulations to assess its applicability in predicting collision probabilities.
Here we aim to construct the joint pdf $p(r,v)$ where $r=(x, y, z)$ is the 3D position vector and $v = (v_x , v_y , v_z)$ is the 3D velocity vector of a space object; we define its entropy as: 
    \begin{equation}
        S[p] = -\int_{\mathbb{R}^3}\int_{\mathbb{R}^3} p(r,v) \ ln\ p(r,v) \ d^3r\ d^3v
    \end{equation}
The pdf must satisfy the normalization constraint:
\begin{equation}
    \int_{\mathbb{R}^3}\int_{\mathbb{R}^3}\ p(r,v) \ d^3r\ d^3v =1
    \label{Eq:1}
\end{equation}
This is the normalization constraint that ensures the total probability to one over all possible states. Moreover this pdf will also ensure the total specific mechanical energy E is conserved, satisfying the energy constraint:
    \begin{equation}
    \int_{\mathbb{R}^3}\int_{\mathbb{R}^3} p(r, v)\ E(r,v) \ d^3r\ d^3v = E_0
    \label{Eq:2}
    \end{equation}
where $E_0$ is the integral constant that ensures the expectation of total energy over the distribution must match this value. In a two-body orbital system, the total specific orbital energy (sum of kinetic and potential energy per unit mass) is:
\begin{equation}
    E(r,v) = \frac{1}{2} |v|^2 - \frac{\mu}{|r|}
\end{equation}
Where $\mu$= GM.
Here we define a new constraint i.e $\mathcal{E}(r,v) = \frac{E(r,v)}{\mu}=\frac{v^2}{\mu} - \frac{1}{r^2} = \mathcal{E}_0 $. In our work, we will use this term as an energy functional throughout the calculations.
Now, to solve this constrained optimization problem, we introduce the Lagrange multipliers $\lambda_1$ and $\lambda_0$ and formulate the Lagrange functional:
\begin{equation}
    \mathcal{L}[p]= -\mathcal{M} +\lambda_o \mathcal{N} +\lambda_1 \mathcal{Q}
\end{equation}
Where $\mathcal{M} = \int_{\mathbb{R}^3}\int_{\mathbb{R}^3} p(r,v) \ ln \ p(r,v) d^3r\ d^3v$, $\mathcal{N} = 1-\int_{\mathbb{R}^3}\int_{\mathbb{R}^3} p(r,v) d^3r\ d^3v$ and $\mathcal{Q} =  \mathcal{E}_0 - \int_{\mathbb{R}^3}\int_{\mathbb{R}^3} p(r,v) \mathcal{E}(r,v) d^3r\ d^3v$. Solving the Lagrange Functional with functional derivative setting $\frac{\delta\mathcal{L}}{\delta p}=0$ we get :
\begin{equation*}
    p(r,v) = e^{[-1 -\lambda_0-\lambda_1 \mathcal{E}(r,v)]}
\end{equation*}
\begin{equation}
 p(r,v)= Ae^{[-\lambda_1 \mathcal{E}(r,v)]}
\end{equation}
Where $A= e^{[-1 -\lambda_0]}$. To derive the value of $A$ using the normalization constraint we substitute the value of p(r,v) in Eq \ref{Eq:1}. Hence, we get:
\begin{equation}
    A \int_{\mathbb{R}^3}\int_{\mathbb{R}^3}e^{-\lambda_1 \mathcal{E}(r,v)} \  d^3r\ d^3v = 1
\end{equation}
Now, we are defining a partition function:
\begin{equation}
Z[\lambda_1] = \int_{\mathbb{R}^3}\int_{\mathbb{R}^3} e^{-\lambda_1 \mathcal{E}(r,v)} \  d^3r\ d^3v
\label{Eq:3}
\end{equation}
Thus it becomes $A=\frac{1}{Z[\lambda_1]}$. Hence we can say that $p(r,v)= \frac{1}{Z[\lambda_1]}  e^{-\lambda_1\mathcal{E}((r,v)}$

Now, we have to compute the value of the Lagrange multiplier $\lambda_1$ with the help of the energy constraint. Substituting the value of $p(r,v)$ in the Eq \ref{Eq:2} we get:
\begin{equation}
    \frac{1}{Z[\lambda_1]}  \int _{\mathbb{R}^3}\int_{\mathbb{R}^3} e^{-\lambda_1 \mathcal{E}((r,v)} \mathcal{E}(r,v) \ d^3r\ d^3v = \mathcal{E}_0
    \label{Eq:4}
\end{equation}
Differentiating the Eq \ref{Eq:3} with respect to $\lambda_1$ we get : 
\begin{equation}
    \frac{d Z[\lambda_1]}{d \lambda_1} = -\int _{\mathbb{R}_3}\int_{\mathbb{R}_3} \mathcal{E}(r,v) \ e^ {-\lambda_1 \mathcal{E}(r,v)} d^3r \ d^3v
\end{equation}
Therefore Eq \ref{Eq:4} becomes 
\begin{equation}
   \mathcal{E}_0 = - \frac{1}{Z(\lambda_1)} \ \frac{dZ[\lambda]}{d \lambda_1} \implies \mathcal{E}_0 = - \frac{d \ ln\ Z[\lambda]}{d \lambda_1}
    \label{Eq:10}
\end{equation}
Putting the values of $E(r,v)$ in Eq \ref{Eq:3} we get: 
\begin{equation}
    Z(\lambda_1) = \int_{\mathbb{R}^3}\int_{\mathbb{R}^3} e^{-\lambda_1( \frac{1}{2\mu} |v|^2-\frac{1}{|r|})} d^3r d^3v 
    \label{Eq:5}
\end{equation}

For the betterment of understanding and calculation, if we break the integral into two parts we get the following:
  \begin{equation}
    I_v= {\int\int\int} e^{(\frac{-\lambda_1 |v|^2}{2\mu})}\ dv_x dv_y dv_z
     \label{Eq:6}
  \end{equation}
  \begin{equation}
    I_r= {\int\int\int} e^{(\frac{\lambda_1}{|r|})}\ dr_x dr_y dr_z
     \label{Eq:7}
  \end{equation}

Converting the velocity integral into a spherical coordinate and solving the first integral in equation \ref{Eq:6} we get $I_v = \left(\frac{2\pi}{\lambda_1}\right)^{\frac{3}{2}} \sqrt{\mu}$.

Similarly, while equating the position integral in equation \ref{Eq:7}, we need to convert it into the spherical coordinate system. In spherical coordinate system the equation becomes: 

\begin{equation}
    I_r= \int_{r_1}^{r_2} \int_{0}^{\pi} \int_{0}^{2\pi} e^{(\frac{\lambda_1}{\rho})} \rho ^2\ sin\theta \  d\phi d\theta d\rho
\end{equation}
 where $r_1 > 0$ and $r_2 < \infty$. Solving the first two angular integrals, we get:
 \begin{equation}
     \int_{0}^{\pi} \int_{0}^{2\pi}  sin\theta \  d\phi d\theta = 4\pi
 \end{equation}

Now using the exponential series expansion $e^{\left(\frac{\lambda_1}{\rho}\right)}$ can be expanded and the integral becomes:
 \begin{equation}
     4\pi \int_{r_1}^{r_2} \rho ^2\left[ 1+\frac{\lambda_1}{\rho}+ \frac{1}{2!}\left(\frac{\lambda_1}{\rho}\right)^2+\frac{1}{3!}\left(\frac{\lambda_1}{\rho}\right)^3+...\right] d\rho
     \label{Eq:8}
 \end{equation}

Now approximating the equation \ref{Eq:8} upto second order we get: 
\begin{equation}
    I_r \approx 4\pi \int_{r_1}^{r_2} \rho^2 \left[1+  \frac{\lambda_1}{\rho}+ \frac{1}{2!}\left(\frac{\lambda_1}{\rho}\right)^2\right] d\rho
\end{equation}
Finally the position integral becomes: $I_r \approx 2 \pi (r_2 - r_1) \left[\frac{2}{3} (r_2^2 + r_2 r_1 + r_1^2) + \lambda_1 (1+r_2+r_1)\right]$
and the value of $Z(\lambda_1)$ becomes:
\begin{align}
    Z[\lambda_1]=
    \left(\frac{2\pi}{\lambda_1}\right)^{(\frac{3}{2})}\sqrt{\mu} \ 2\pi \ [ \frac{2}{3}(r_2^2 + r_1r_2 + r_1^2)+ \lambda_1 (1+r_2+r_1)]
    \label{Eq:9}
\end{align}

Now if we simplify the structure of $Z[\lambda_1]$ then it becomes $C_1 \lambda_1^{-(\frac{3}{2})}+ C_2 \lambda_1^{-(\frac{1}{2})}$ where the value of $C_1= \frac{2}{3}(2 \pi)^{\frac{5}{2}}\sqrt{\mu}(r_2-r_1)(r_2^2 + r_1r_2 + r_1^2)$ and $C_2 = \frac{2}{3}(2 \pi)^{\frac{5}{2}}\sqrt{\mu}(r_2-r_1)(1+r_2+r_1)$. Now differentiating $Z[\lambda_1]$ with respect to $\lambda_1$ we get $\frac{d[Z[\lambda_1]]}{d\lambda_1} = -(\frac{3}{2}) C_1 \lambda_1^{-(\frac{5}{2})}+ -(\frac{1}{2}) C_2 \lambda_1^{-(\frac{3}{2})}$. Here we assume $-(\frac{3}{2}) C_1 = A_1$ and $-(\frac{1}{2}) C_2 = A_2$ which leads to the value of $\lambda_1$ to be:
\begin{equation}
   \lambda_1 = \frac{2A_1}{-(A_2 - \mathcal{E}_0 C_1) \pm \sqrt{(A_2 - \mathcal{E}_0)^2+ 4A_1\mathcal{E}_0C_1}}
\end{equation}
Now we look into the square root term of the denominator of $\lambda_1$. We can see that the value of $A_2$ is an extremely large one. Hence we can say $(A_2 - \mathcal{E}_0)^2 \approx A_2^2$. Similarly, the term $ 4A_1\mathcal{E}_0C_1$ is also very small to $A_2^2$. Therefore, we can neglect that term, too. Thus it becomes: $\sqrt{(A_2 - \mathcal{E}_0)^2+ 4A_1\mathcal{E}_0C_1} \approx \sqrt{A_2^2} = A_2$

Hence finally the value of $\lambda_1$ can be written as:
\begin{equation}
    \lambda_1 \approx \frac{2A_1}{-A_2 + \mathcal{E}_0 C_1\pm A_2}
\end{equation}
We are getting two values of $\lambda_1$ here i.e. $\lambda_{11} \approx \frac{2A_1}{\mathcal{E}_0 C_1}$ and  $\lambda_{12} \approx \frac{2A_1}{\mathcal{E}_0 C_1 - 2A_2}$.

While validating the PDF $\lambda_{11}$ is producing an extremely small value. Hence $\lambda_{12}$ becomes an appropriate one providing a much more feasible PDF value.

Thus the final PDF is in the form $p(r,v)= \frac{1}{Z[\lambda_1]}  e^{-\lambda_1\mathcal{E}((r,v)}$ where the values of $Z[\lambda_1]$, $\lambda_1$, $\mathcal{E}_0$ \& $\mathcal{E}(r,v)$ have already been deduced and mentioned above.

\section{Results and Discussion}
Our study successfully derived a 6-dimensional PDF to model the uncertainty in the position and velocity of space objects in Low Earth Orbit (LEO). By applying the MaxEnt, we formulated a distribution that adheres to energy conservation while capturing the inherent uncertainties in orbital dynamics. The resulting PDF provides a robust framework for predicting collision risks and assessing orbital uncertainties without relying on restrictive assumptions about the underlying dynamics.

A key challenge in this work was handling the exponential terms in the PDF, which could lead to numerical instabilities when the Lagrange multiplier $\lambda_1$ becomes large. To address this, we employed approximations in some places . This approach ensured that the PDF remained well-behaved and computationally tractable, even for high-energy constraints.

\begin{figure}[ht]
    \centering
    \includegraphics[width=0.4\linewidth]{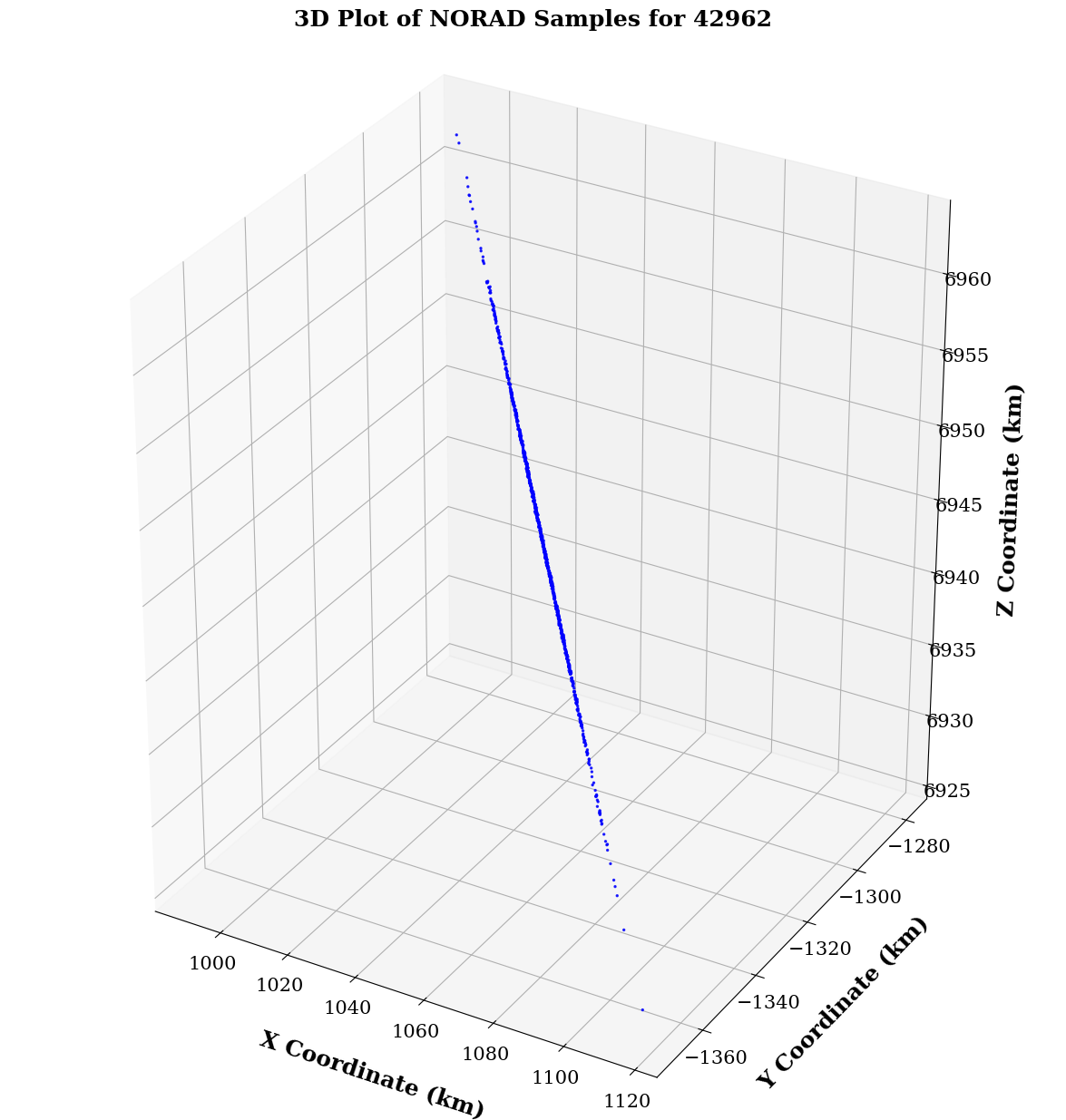}
    \caption{Distribution of Samples in 3 dimension}
    \label{fig:Distribution of Samples}
\end{figure}
\begin{figure}[ht]
    \centering
    \includegraphics[width=0.4\linewidth]{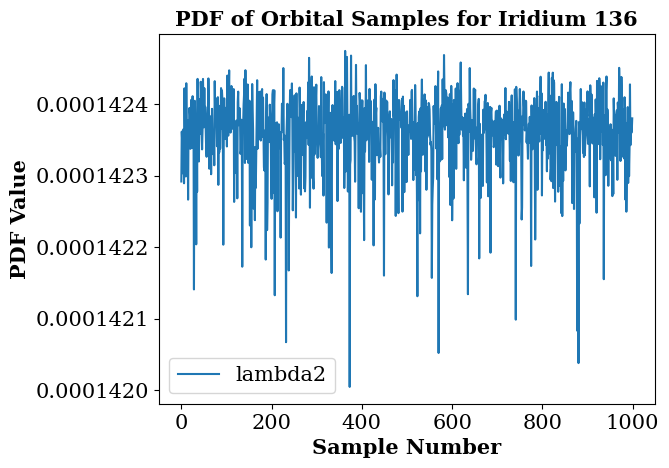}
    \caption{PDF for $\lambda_{12}$}
    \label{fig:PDF for lambda12}
\end{figure}
\begin{figure}[ht]
    \centering
    \includegraphics[width=0.4\linewidth]{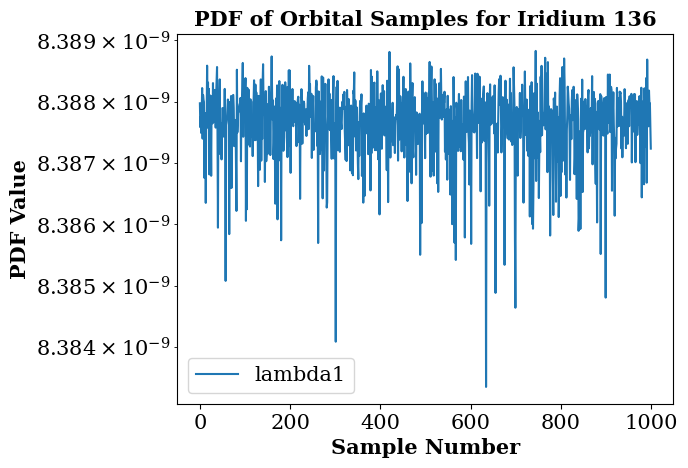}
    \caption{PDF for $\lambda_{11}$}
    \label{fig:PDF for lambda11}
\end{figure}

To test the PDF numerically, we fetched TLE data of Iridium 136 (NORAD 42962) from spacetrack.org\href{spacetrack.org}. We generated 1000 samples of around the position mean (fig:\ref{fig:Distribution of Samples}) and tested the PDF for both the values of $\lambda_1$. As a result we found that for $\lambda_{11}$ the PDF is almost converging to zero (fig \ref{fig:PDF for lambda11}, which is not expected. But the value of the PDF for $\lambda_{12}$ was quite decent (fig: \ref{fig:PDF for lambda12}).

\begin{figure}[ht]
    \centering
    \includegraphics[width=0.5\linewidth]{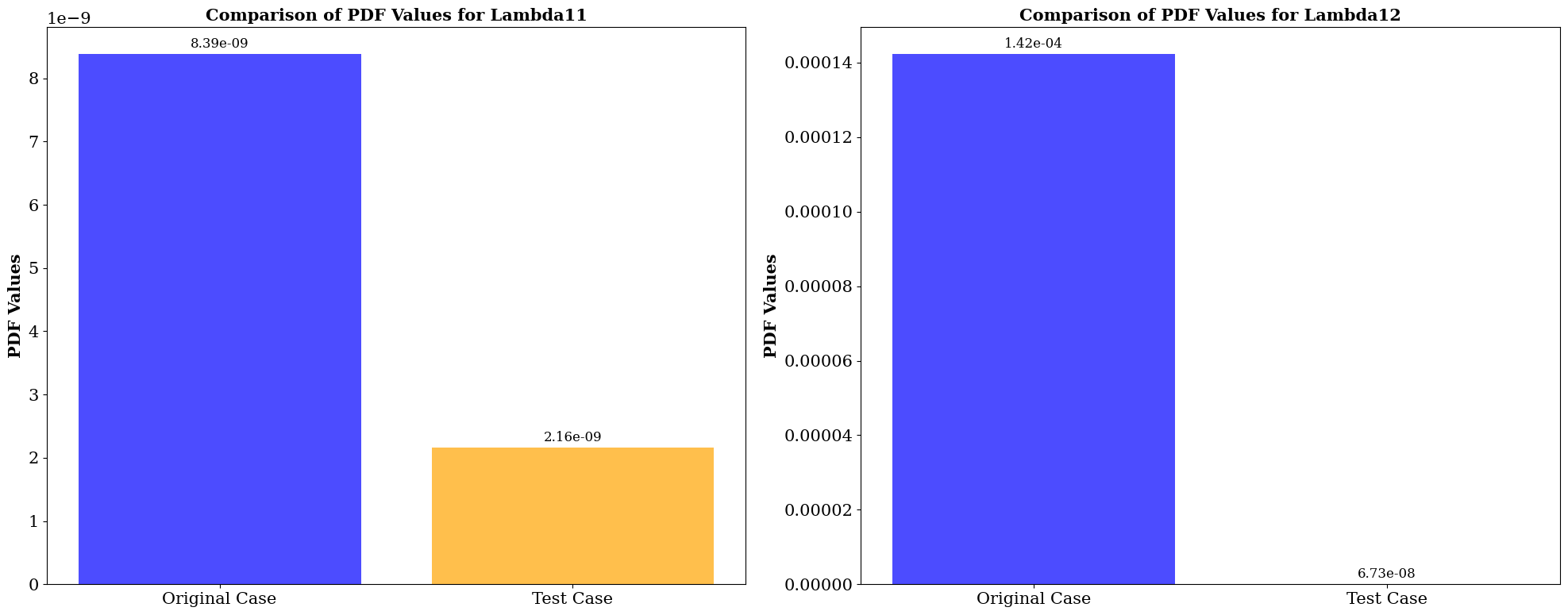}
    \caption{Validation of the PDF for $\lambda_{11}$ \& $\lambda_{12}$}
    \label{fig:Validation}
\end{figure}

Finally, we also tested the accuracy of the PDF in fig: \ref{fig:Validation}, adding randomly large noise to the position mean of the TLE data, which led to the PDF values being almost zero. Hence it validates our PDF through numarical testing. 

Our analysis also highlighted the importance of selecting appropriate constraints while maximizing entropy. By carefully balancing the energy conservation requirement with the need for a physically meaningful distribution, we obtained a PDF that accurately reflects the statistical behaviour of space objects. Numerical validation confirmed that the derived distribution effectively captures the expected energy fluctuations while maintaining stability across different orbital regimes.

\section{Conclusion}
This study presents a novel approach to modelling orbital uncertainty using maximum entropy principles, offering a flexible and physics-based alternative to traditional Gaussian approximations. The derived six-dimensional PDF provides a foundation for advanced collision risk assessment and uncertainty propagation in LEO, where the growing number of satellites and debris demands improved predictive tools.
Future research will focus on refining approximations as well as integrating real-time tracking data to enhance space traffic management, ultimately promoting safer and more sustainable space operations. By bridging the gap between statistical mechanics and orbital dynamics, this research will contribute to the betterment of space situational awareness in such an increasingly congested orbital environment.

\bibliographystyle{unsrt}  


\end{document}